# Lasing without inversion in a closed four-level system

Li-Feng Cai

*Department of Applied Physics, National University of Defence Technology, Changsha, 410073，People's Republic of aChina*

We show lasing without population inversion in a closed four-level atomic system which is no longer incoherently pumped on the transition at which lasing occurs. The four-level system can overcome some limits in the typical lambda-type and V-type three-level systems, furthermore it is possible to achieve lasing even in the absence of the inversion condition for the thermal radiation fields or of the inversion of spontaneous decay rate. We find that when the detuning of probe field is equal to that of coherent coupling field, there is a peak value along with the increase of the detuning. To explain it, we bring forward that there are adiabatic and nonadiabatic stimulated Raman processes when a three-level atomic system is interacted with two coherent fields, so there are two mechanisms influencing the change of probe field. We believe that it can explain more general lasing without inversion processes because of quantum interference induced by coherent fields.

## Ⅰ.INTRODUTION

Since it is firstly presented in 1989[1,2], the studies of lasing without inversion in multilevel atomic systems have developed for almost twenty years. Many schemes for lasing without inversion have been proposed[3-6],and their physics mechanisms have been discussed[7-10].Experimental observations of inversionless gain and lasing have been reported[11,12] .In certain conditions, lasing without inversion in multilevel atomic systems can be viewed as a generalization of earlier results observed in studies of coherently driven two-level systems[13].But generally, multilevel atomic systems, as the typical Λ-type and V-type three-level systems, lasing without inversion because of quantum interference between atomic levels induced by coherent fields, exhibits novel light amplification phenomena different from the two-level systems. For example, the multilevel systems can crease laser gain at frequencies far removed from that of the coherent coupling fields, thus provide possibilities for the generation of coherent radiation in the short-wave length regime., furthermore there are novel statistical properties, such as reduced laser linewidth[14] and so on.

Generally speaking, LWI can be classified into two categories: first, LWI in any state basis; second, LWI in the bare atomic states, but with inversion in a hidden-state basis, such as dressed states. As is shown in the ref[4], when the coupling field is detuned from the atomic transition, there are a inversion in the dressed states which is accounted for the light amplification. But it is limited in the condition that the coupling field is far stronger than the probe field. When the probe field is comparable to the coupling field, there is still lasing without inversion, which cannot be attributed to the inversion in the dressed states. In fact, we find that even in detuning there is still lasing without inversion in any state basis. This bring us to find out a unified explanation which can explain the light amplification without inversion in any state basis, also explain that with inversion in dressed state basis under the given condition.

The process that a three-level atomic system interacted with two coherent fields is substantially the stimulated Raman processes. We think that because of quantum interference between atomic levels induced by coherent fields the stimulated Raman processes is divided between adiabatic one and nonadiabatic one, so there are two mechanisms influencing the change of probe field. From which, we propose and explain a closed four-level atomic model for lasing without inversion.

Λ-type and V-type three-level systems shown in fig.1[3,4] are two typical models for lasing without inversion. For gain on the transition at which lasing occurs, the condition in three-level systems is that the spontaneous decay rate from state $|3\rangle$ to $|2\rangle$ in Λ-type ($|2\rangle$ to $|1\rangle$ in V-type) exceeds that from state $|3\rangle$ to $|1\rangle$ in Λ-type ($|3\rangle$ to $|1\rangle$ in V-type), and the average number of thermal photons per mode in the $|3\rangle \leftrightarrow |1\rangle$ channel in Λ-type ($|3\rangle \leftrightarrow |1\rangle$ channel in V-type) exceeds that in the $|3\rangle \leftrightarrow |2\rangle$ channel in Λ-type ($|2\rangle \leftrightarrow |1\rangle$ channel in V-type). Both models require that the spontaneous decay rate between lower energy levels exceeds that between higher energy levels and the average number of thermal photons per mode between higher energy-level exceeds that between lower energy levels. According to the $w^3$ scaling of Einstein A coefficient, and the average number of thermal photons per mode $\bar{n} = 1/(e^{\hbar w/kT} - 1)$, it isn't easy to get the two conditions. Furthermore both models require the incoherent pumping between the transition at which lasing occurs. In the generation of higher-frequency lasers via LWI, the incoherent pumping is still the mail obstacle. In this communication, we propose a closed four-level system which can overcome some limits in the typical Λ-type and V-type three-level systems. It will be shown that no both inversion conditions in the spontaneous decay rate and the average number of thermal photons per mode must be satisfied, but only either one of them is satisfied. Furthermore there is no need to be incoherently pumped on the transition at which lasing occurs, but only another lower frequency transition. Then we discuss the LWI in detuning. When the probe field is far weaker than the coherent coupling field, considering the Auterler-Townes dressed states, we get the origin of lasing without inversion from the deduced Hamiltonian of the atomic system. But when the probe field is not very weak comparable with the coupling field, we find that when the detuning of probe field is equal to that of coherent coupling field, there is a peak value along with the increase of the detuning. To explain it, we bring forward that there are two mechanisms influencing the change of probe field. Just because of the competition between them, LWI is generated. We believe that it can explain more general lasing without inversion processes because of quantum interference induced by coherent fields.

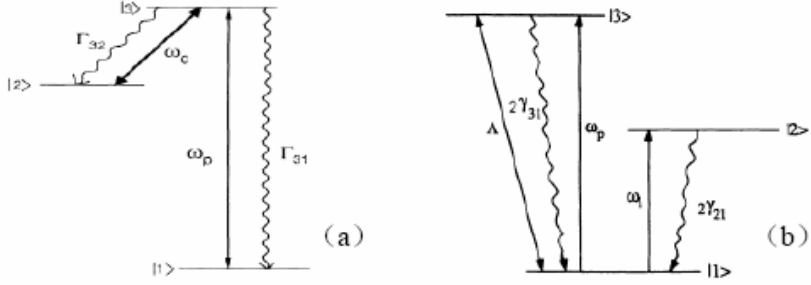

FIG.1.Λ-type（a） 和 V-type（b） three-level system for lasing without inversion

## II. EQUATION OF MOTION

We consider a four-level system as illustrated in fig.2. The transition $|a\rangle \leftrightarrow |c\rangle$ of frequency $w_{ac}$ is driven by a strong coherent field of frequency $w_1$ with Rabi frequency $\Omega$. A weak, coherent probe field of frequency $w_2$ with Rabi frequency g is applied to the transition $|a\rangle \leftrightarrow |b\rangle$. The transition $|c\rangle \leftrightarrow |d\rangle$ of frequency $w_{cd}$ is incoherently pumped with a rate $R_{cd}$ and the transition $|b\rangle \leftrightarrow |d\rangle$ of frequency $w_{bd}$ is incoherently pumped with a rate $R_{bd}$. $\gamma_{ij}$ is the spontaneous decay rate from $|i\rangle$ to $|j\rangle$ ($i, j = a, b, c, d$). The transitions $|c\rangle \leftrightarrow |b\rangle$ and $|a\rangle \leftrightarrow |d\rangle$ are thus dipole forbidden. The semi-classical Hamiltonian under rotating-wave approximation is: $H = H_0' + H_0'' + H_1$, where

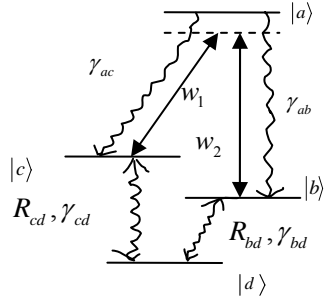

FIG.2. Four-level system for lasing without inversion

$$H_0' = w_a |a\rangle\langle a| + (w_a - w_2)|b\rangle\langle b| + (w_a - w_1)|c\rangle\langle c| + w_d |d\rangle\langle d|$$
$$H_0'' = \Delta_2 |b\rangle\langle b| + \Delta_1 |c\rangle\langle c| \qquad (1)$$
$$H_1 = -\frac{1}{2}\left(\Omega |a\rangle\langle c| e^{-iw_1 t} + g |a\rangle\langle b| e^{-iw_2 t}\right) + H.c$$

Here $\Delta_1 = w_1 - w_a + w_c$, $\Delta_2 = w_2 - w_a + w_b$. Without loss of generality, $\Omega$ and g are chosen to be positive real number by adding a global phase to the states $|b\rangle$ and $|c\rangle$, respectively. Following the method used with the interaction representation, we attribute the change of operators to $H_0'$, then the change of states is followed the Hamiltonian

$$H' = e^{iH_0' t}\left(H_0'' + H_1\right)e^{-iH_0' t} = \Delta_2 |b\rangle\langle b| + \Delta_1 |c\rangle\langle c| - \frac{1}{2}\left[\left(\Omega |a\rangle\langle c| + g |a\rangle\langle b|\right) + H.c\right] \qquad (2)$$

Where $\hbar$ is set to 1. Using the quantum theory of damping [15], the master equation can be written as

$$\frac{d\rho}{dt} = -i[H', \rho] + L\rho \qquad (3)$$

where the damping and incoherent pumping term $L\rho$ is

$$L\rho = -\sum_{j=b,c} \frac{\gamma_{aj}}{2}\left(\sigma_{aj}\sigma_{ja}\rho - 2\sigma_{ja}\rho\sigma_{aj} + \rho\sigma_{aj}\sigma_{ja}\right) - \sum_{j=b,c} \frac{R_{jd}+\gamma_{jd}}{2}\left(\sigma_{jd}\sigma_{dj}\rho - 2\sigma_{dj}\rho\sigma_{jd} + \rho\sigma_{jd}\sigma_{dj}\right)$$
$$-\sum_{j=b,c} \frac{R_{jd}}{2}\left(\sigma_{dj}\sigma_{jd}\rho - 2\sigma_{jd}\rho\sigma_{dj} + \rho\sigma_{jd}\sigma_{dj}\right) \quad (4)$$

Here $\sigma_{ij} = |i\rangle\langle j| \, (i,j=a,b,c,d)$ are the atomic raising or lowering operators. Incoherent pumping rate $R_{jd} = n_{jd}\gamma_{jd}\,(j=b,c)$, $n_{jd}$ are the average number of thermal photons in the $|j\rangle \leftrightarrow |d\rangle$ channel. The density matrix $\rho$ in Equations (3) (4) is determined by $H'$ which is not the Hamiltonian in the interaction representation. But we can convert (3) (4) to the interaction representation by $H_0''$, then we will get just the same form as (3) (4), so (3) (4) is correct. From (3) (4), we have

$$\rho_{aa} + \rho_{bb} + \rho_{cc} + \rho_{dd} = 1$$
$$\dot\rho_{bb} = \gamma_{ab}\rho_{aa} + \frac{i}{2}g(\rho_{ab}-\rho_{ba}) + R_{bd}(\rho_{dd}-\rho_{bb}) - \gamma_{bd}\rho_{bb}$$
$$\dot\rho_{cc} = \gamma_{ac}\rho_{aa} + \frac{i}{2}\Omega(\rho_{ac}-\rho_{ca}) + R_{cd}(\rho_{dd}-\rho_{cc}) - \gamma_{cd}\rho_{cc}$$
$$\dot\rho_{dd} = -R_{cd}(\rho_{dd}-\rho_{cc}) - R_{bd}(\rho_{dd}-\rho_{bb}) + \gamma_{bd}\rho_{bb} + \gamma_{cd}\rho_{cc} \quad (5)$$
$$\dot\rho_{bc} = -\frac{1}{2}\left[R_{cd}+R_{bd}+\gamma_{cd}+\gamma_{bd}+i2(\Delta_2-\Delta_1)\right]\rho_{bc} - i\frac{\Omega}{2}\rho_{ba} + i\frac{g}{2}\rho_{ac}$$
$$\dot\rho_{ba} = -\frac{1}{2}(R_{bd}+\gamma_{bd}+\gamma_{ab}+\gamma_{ac}+i2\Delta_2)\rho_{ba} + i\frac{g}{2}(\rho_{aa}-\rho_{bb}) - i\frac{\Omega}{2}\rho_{bc}$$
$$\dot\rho_{ca} = -\frac{1}{2}(R_{cd}+\gamma_{cd}+\gamma_{ab}+\gamma_{ac}+i2\Delta_1)\rho_{ca} + i\frac{\Omega}{2}(\rho_{aa}-\rho_{cc}) - i\frac{g}{2}\rho_{cb}$$

In the following, according to (5), we discuss the change of the probe field coupled to the transition $|a\rangle \leftrightarrow |b\rangle$. According to Maxwell equation for slowly varying field functions, the gain coefficient for the probe field is proportional to the negative imaginary part of positive frequency part of atomic polarization, i.e., $\dot E \propto -\mathrm{Im}\, p$. The positive frequency part of average value of dipole moment operator is $p = \frac{g}{E}\rho_{ab}$, therefore $\dot E \propto \mathrm{Im}\,\rho_{ba}$. In following discusses, we seek a steady-state solution and set the derivatives in equation (5) equal to zero. If $\mathrm{Im}\,\rho_{ba} > 0$, the system exhibits gain for the probe field. At first, $\Delta_1 = \Delta_2 = 0$ is considered; then we discuss the instance that $\Delta_1$ and $\Delta_2$ aren't equal to zero.

### III. GAIN AND POPULATION DISTRIBUTION ON RESONANCE

When $\Delta_1 = \Delta_2 = 0$, and for the sake of simplicity, we suppose that only the transition $|c\rangle \leftrightarrow |d\rangle$ is incoherently pumped, i.e., $R_{bd} = 0$ we seek the steady-state solution of $\mathrm{Im}\,\rho_{ba}$ and population distribution. Now, $\mathrm{Re}\,\rho_{ba} = \mathrm{Re}\,\rho_{ca} = \mathrm{Im}\,\rho_{bc} = 0$. In the limit

of $\Omega \gg g, \gamma_{i,j}, R_{cd}$, from (5) it can to be obtained that:

$$\text{Im}\,\rho_{ba} = \frac{gR_{cd}}{\Omega^2} \frac{(\gamma_{cd}+\gamma_{bd}+R_{cd})(\gamma_{bd}-\gamma_{ab})+\gamma_{bd}(\gamma_{ac}+\gamma_{ab})}{(2R_{cd}+\gamma_{cd})\gamma_{bd}+R_{cd}(\gamma_{bd}+\gamma_{ab})+\gamma_{bd}\gamma_{ab}} \tag{6}$$

The condition for the amplification of weak probe field is

$$\gamma_{bd} > \gamma_{ab} \quad \text{or} \quad 0 < \gamma_{ab}-\gamma_{bd} < \frac{\gamma_{bd}(\gamma_{ac}+\gamma_{ab})}{\gamma_{cd}+\gamma_{bd}+R_{cd}} \tag{7}$$

The population distribution is

$$\rho_{aa} = \rho_{cc} = \frac{\gamma_{bd}R_{cd}}{(2R_{cd}+\gamma_{cd})\gamma_{bd}+R_{cd}(\gamma_{bd}+\gamma_{ab})+\gamma_{bd}\gamma_{ab}} = \frac{\gamma_{bd}}{\gamma_{ab}}\rho_{bb} = \frac{R_{cd}}{R_{cd}+\gamma_{cd}+\gamma_{ab}}\rho_{dd} \tag{8}$$

When $\gamma_{ab}-\gamma_{bd} > 0$, then $\rho_{aa} \leq \rho_{cc} < \rho_{bb}$, there is no population inversion in the bare-state basis. The other meaningful state basis is the dressed-state basis consist of states $|b\rangle, |+\rangle$ and $|-\rangle$. For a resonant coupling laser, the dressed states formed by strong coupling laser with Rabi frequency $\Omega$ is (see (9)) $|\pm\rangle = \frac{1}{\sqrt{2}}(|a\rangle \pm |c\rangle)$. from (8) and $\text{Re}\,\rho_{ca} = 0$, the population distribution in the dressed states is $\rho_{++} = \rho_{--} = \frac{\rho_{aa}+\rho_{cc}}{2} < \rho_{bb}$, no population inversion in the dressed states either. Because the frequency between the transition $|a\rangle$ and $|b\rangle$ is far larger than that between $|b\rangle$ and $|d\rangle$, generally $\gamma_{ab}-\gamma_{bd} > 0$. So only the condition $0 < \gamma_{ab}-\gamma_{bd} < \frac{\gamma_{bd}(\gamma_{ac}+\gamma_{ab})}{\gamma_{cd}+\gamma_{bd}+R_{cd}}$ is met, the atomic system will exhibit gain without population inversion. Deferent from that in $\Lambda$-type and V-type three-level systems, here it is no longer required to satisfy both inversion conditions i.e. $n_{cd} = \frac{R_{cd}}{\gamma_{cd}} > n_{bd} = \frac{R_{bd}}{\gamma_{bd}}$ and $\gamma_{ac} > \gamma_{ab}$, but only $n_{cd} = \frac{R_{cd}}{\gamma_{cd}} > n_{bd} = \frac{R_{bd}}{\gamma_{bd}}$, namely the inversion in the average number of thermal photons. On the other hand, if only $\gamma_{ac} > \gamma_{ab}$, but $n_{cd} < n_{bd}$, it can also be shown lasing without inversion. When $R_{bd}$ is no longer zero, the analytical expression of $\text{Im}\,\rho_{ba}$ is complicated. But with numerically calculation, we find the example for lasing without inversion when $\gamma_{ac} > \gamma_{ab}$ and $n_{cd} < n_{bd}$. The response of $\text{Im}\,\rho_{ba}$ vs $\gamma_{ac}$ in units of the decay $\gamma_{ab}$ is plotted in Fig3(a) with chosen parameters: $\Omega = 10\gamma_{ab}$, $g = \gamma_{ab}$, $\gamma_{cd} = \gamma_{bd} = 0.5\gamma_{ab}$, $n_{bd} = 2$, and $n_{cd} = 1$. For $n_{bd} = 2, n_{cd} = 1$, namely without the inversion in the average number of thermal photons, when $\gamma_{ac}/\gamma_{ab}$ is about 10~15, $\mathrm{Im}\rho_{ba} > 0$, hence the probe field is amplified. Checking the population distribution, we set $\gamma_{ac}/\gamma_{ab} = 12$, it is obtained that $\rho_{aa} = 0.0573$, $\rho_{cc} = 0.1643, \rho_{bb} = 0.3344, \rho_{dd} = 0.4440$, $\rho_{aa} < \rho_{cc} < \rho_{bb} < \rho_{dd}$, namely there is no population inversion. So we have shown that the system can exhibit lasing without inversion even without the inversion in the average number of thermal photons. We will explain these phenomena in the sec. Ⅴ.

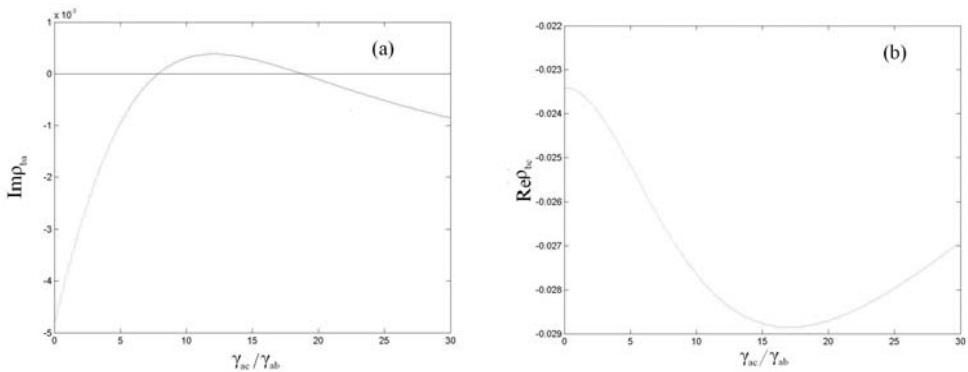

FIG.3. (a) For without the inversion in the average number of thermal photons, the response $\text{Im}\,\rho_{ba}$ as functions of the normalized decay $\gamma_{ac}$. $\gamma_{ac}/\gamma_{ab}$ is about 10~15, $\text{Im}\,\rho_{ba} > 0$ (b) The reponse of $\text{Re}\,\rho_{bc}$ vs $\gamma_{ac}/\gamma_{ab}$. Values of the parameters are $\Omega = 10\gamma_{ab}$, $g = \gamma_{ab}$, $\gamma_{cd} = \gamma_{bd} = 0.5\gamma_{ab}$, $n_{bd} = 2, n_{cd} = 1$。

## Ⅳ. OFF RESONANCE AND EXPLAINATION

For $\Delta_1$ and $\Delta_2$ are not equal to zero, we calculate numerically the steady-state solution in Eq.(5). At first, we consider the population distribution in bare and dressed state basis. In the limit of $\Omega \gg g$, the strong coherent field with Rabi frequency $\Omega$ couples states $|a\rangle$ and $|c\rangle$, and formed the Autler-Townes dressed stares. Seeking the eigenstates and eigenvalues of $H'$ in (2) relating to states $|a\rangle$ and $|c\rangle$, we can write $H'$ as

$$H' = \Delta_2 |b\rangle\langle b| + \lambda_+ |+\rangle\langle +| + \lambda_- |-\rangle\langle -| + \frac{1}{2}\left[g\left(\sin\theta_+ |+\rangle\langle b| + \cos\theta_+ |-\rangle\langle b|\right) + H.c\right] \quad (9)$$

Where the dressed states is $|\pm\rangle = \sin\theta_\pm |a\rangle + \cos\theta_\pm |c\rangle$, and $\tan\theta_\pm = \frac{-\Omega}{2\lambda_\pm}, \lambda_\pm = \frac{1}{2}(\Delta_1 \mp \sqrt{\Delta_1^2 + \Omega^2})$ is the eigenvalues of $|\pm\rangle$. Again following the method used with the interaction representation, we attribute the change of operators to $\Delta_2 |b\rangle\langle b| + \lambda_+ |+\rangle\langle +| + \lambda_- |-\rangle\langle -|$, then the change of states is followed the Hamiltonian

$$H'' = \frac{1}{2} g \left(\sin\theta_+ |+\rangle\langle b| e^{i(\lambda_+ - \Delta_2)t} + \cos\theta_+ |-\rangle\langle b| e^{i(\lambda_- - \Delta_2)t}\right) + H.c \quad (10)$$

We may choose the value of $\Delta_2$ to make either one of $\lambda_+ - \Delta_2$ and $\lambda_- - \Delta_2$ to equal to zero. Now, we make $\lambda_+ - \Delta_2 = 0$. In the limit of $\Omega \gg g$, we obtain $(\lambda_- - \Delta_2) = \sqrt{\Delta_1^2 + \Omega^2} \gg g$, so we may use the rotating-wave approximation and ignore the counter-rotating terms proportional to $e^{\pm i(\lambda_- - \Delta_2)t}$, (10) is rewritten as

$$H'' = \frac{1}{2}\left(g \sin\theta_+ |+\rangle\langle b| + H.c\right) \quad (11)$$

In this way, $H''$ has been completely converted to the Rabi oscillation of states $|+\rangle$ and $|b\rangle$. It's the same as the theory of lasing with inversion between two-level systems. If the population of state $|+\rangle$ exceeds that of state $|b\rangle$, there is gain for the weak probe field. For $g < \gamma_{ij}$, we may use the time-dependent perturbation theory and the first-order term suffices.

The transition rate for first-order is $\frac{1}{\tau}(|+\rangle \leftrightarrow |b\rangle) \propto |\langle +|H_I|b\rangle|^2$, so the net gain in the transition from $|+\rangle$ to $|b\rangle$ is proportional to $(g\sin\theta_+)^2(\rho_{++} - \rho_{bb})$. The response of $\text{Im}\rho_{ba}$ and $\rho_{++} - \rho_{bb}$ as functions of $\Delta_1$ in units of the decay $\gamma_{bd}$ is plotted in the Fig.(a). Values of chosen parameters are $\Omega = 10\gamma_{bd}, g = \gamma_{bd}, \gamma_{ac} = 2\gamma_{bd}, \gamma_{cd} = 1.5\gamma_{bd}, \gamma_{ab} = 2\gamma_{bd}, R_{cd} = \gamma_{bd}, R_{bd} = 0$, $\Delta_2 = (\Delta_1 - \sqrt{\Delta_1^2 + \Omega^2})/2$. In Fig.4(a), about $\Delta_1 < -20\gamma_{bd}$, the probe field is amplified and emerges a peak. While $\Delta_1$ increases from zero, the probe field is attenuated and the attenuation is gradually diminished. Then we see the Fig.4(b), for $\Delta_1 < -20\gamma_{bd}$, then $\rho_{++} - \rho_{bb} > 0$, and

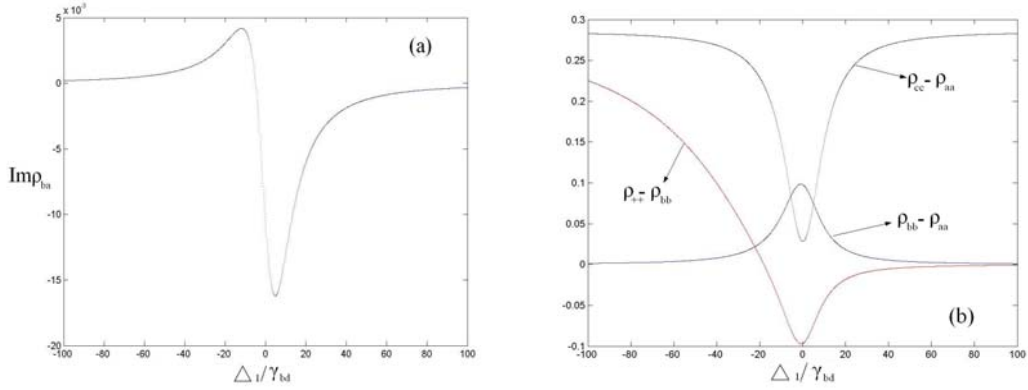

FIG.4. (a) Calculated response of $\text{Im}\rho_{ba}$ as functions of $\Delta_1$. While $\Delta_1 < 0$, probe field is amplified, and there is a peak at $0 \sim \pm 20\gamma_{bd}$. (b) Response of $\rho_{++} - \rho_{bb}$ and population inversions as functions of $\Delta_1$. There is not inversion in the bare states. For $\Delta_1 < 0$, there exist $\rho_{++} - \rho_{bb} > 0$. While $\Delta_1 > 0$, $\rho_{++} - \rho_{bb}$ gradually decreases. The chosen parameters are $\Omega = 10\gamma_{bd}, g = \gamma_{bd}, \gamma_{ac} = 2\gamma_{bd}, \gamma_{cd} = 1.5\gamma_{bd}$, $\gamma_{ab} = 2\gamma_{bd}, R_{cd} = \gamma_{bd}, R_{bd} = 0, \Delta_2 = (\Delta_1 - \sqrt{\Delta_1^2 + \Omega^2})/2$.

$\rho_{bb} - \rho_{aa} > 0, \rho_{cc} - \rho_{aa} > 0$, there is inversion in the dressed states but there is no inversion in the bare stares. During which, though $\rho_{++} - \rho_{bb}$ gradually increases, but $\sin\theta_+$ gradually decreases, so there occurs a peak of gain for probe field which is determined by $(g\sin\theta_+)^2(\rho_{++} - \rho_{bb})$. For $\Delta_1 > 0$, $\rho_{++} - \rho_{bb} < 0$, there is not inversion in the dressed states but $\rho_{++} - \rho_{bb}$ gradually decreases, so the probe field is attenuated and the attenuation is gradually diminished. These are completely consistent with the expectation according to $H''$ in Eq.(11). Nevertheless, it is also shown in Fig.4(a) that the probe field is not always be attenuated for $\Delta_1$ from 0 to $-20\gamma_{bd}$, while there is no inversion in any state basis, either the bare state basis or the dressed state basis. It is no longer consistent with Eq.(11). It is because that Eq.(11) is got in the condition that $\Omega \gg g$, but in Fig4, $\Omega = 10g$, there is only difference of one order. When we continue increasing the probe field, the inconsistency will be more evident. It tells that when the difference between $\Omega$ and $g$ is not very large the Autler-Townes dressed state basis dose not suffice.

We find, as is shown in Fig.5(a), for $\Delta_1 = \Delta_2 = \Delta$, the probe field is always amplified in a wide range of $\Delta$ ($>\Omega, g, \gamma_{ij}, R_{cd}$), furthermore there is a peak along with the increase of absolute value of $\Delta$. The peak value is about one order higher than that in resonance. For the population distribution, there is not population inversion, even $\rho_{cc} < \rho_{bb}$ between dot A and dot B. To explain it, we convert the Hamiltonian $H'$ in Eq.(2) into.

$$H' = -\Delta|a\rangle\langle a| - \frac{1}{2}\left[(\Omega|a\rangle\langle c| + g|a\rangle\langle b|) + H.c\right] - \Delta|d\rangle\langle d| \quad (12)$$

Here we have set $\Delta_1 = \Delta_2 = \Delta$ and employed the completeness relation $\sum_{i=a,b,c,d}|i\rangle\langle i| = 1$, and ignored the constant energy term $\Delta$. Now, we seek the eigenstates and eigenvalues of $H'$ in Eq.(12), we obtain

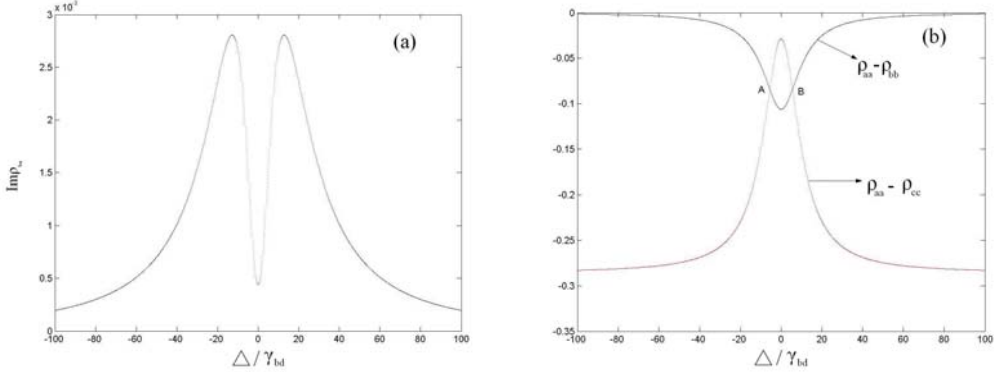

FIG.5. (a) For $\Delta_1 = \Delta_2 = \Delta$, the response of $\mathrm{Im}\rho_{ba}$ as functions of normalized detuning $\Delta/\gamma_{bd}$. Probe field is always amplified in a wide range of $\Delta$ ($>\Omega, g, \gamma_{ij}, R_{cd}$), and there is a peak in the range of $\Delta \approx 0 \sim \pm 20$.

(b) For $\Delta_1 = \Delta_2 = \Delta$, a plot of $\rho_{aa} - \rho_{bb}$ and $\rho_{aa} - \rho_{cc}$ versus normalized detuning $\Delta/\gamma_{bd}$. There isn't population inversion, even $\rho_{cc} < \rho_{bb}$ between dot A and dot B. Besides $\Delta_1 = \Delta_2 = \Delta$, the chosen parameters are the same as that in fig.4.

$$\begin{aligned}
|E_+\rangle &= \cos\phi|a\rangle + \sin\phi\sin\theta|b\rangle + \sin\phi\cos\theta|c\rangle, E_+ = \frac{1}{2}(\Delta + \sqrt{\Delta^2 + G^2}) \\
|E_-\rangle &= -\sin\phi|a\rangle + \cos\phi\sin\theta|b\rangle + \cos\phi\cos\theta|c\rangle, E_- = \frac{1}{2}(\Delta - \sqrt{\Delta^2 + G^2}) \quad (13) \\
|E_0\rangle &= \cos\theta|b\rangle - \sin\theta|c\rangle, \quad E_0 = 0
\end{aligned}$$

Where $\tan\theta = \frac{g}{\Omega}, \tan\phi = \frac{-G}{2E_+} = \frac{2E_-}{G}, G = \sqrt{g^2 + \Omega^2}$. $|E_0\rangle$ is called dark state, if atom is initially prepared in $|E_0\rangle$, there is no absorption even in the presence of the field. From (13) $H'$ can be rewritten as

$$H' = -G/2\left[(\tan\phi\sin^2\phi + \cot\phi\cos^2\phi)|a\rangle\langle a| + (|\uparrow\rangle\langle a| + |a\rangle\langle\uparrow|)\right] - \Delta|d\rangle\langle d| \quad (14)$$

Where $|\uparrow\rangle = \sin\theta|b\rangle + \cos\theta|c\rangle$. Now $H'$ has been converted to the Rabi oscillation of states $|+\rangle$ and $|b\rangle$ as well. If $\rho_{aa} > \rho_{\uparrow\uparrow}$, then the probe field will be amplified due to population inversion in states $|\uparrow\rangle$ and $|a\rangle$. But, in fact, as shown in Fig.6(b), there are not inversion in them, furthermore we cannot search for any other meaningful inversion(see Fig.6(a)). Therefore, the physics behind the lasing without inversion can not be explained only by searching for some kind of hidden inversion. This brings us to think over whether there is another novel mechanism account for the lasing without inversion?

We assume the atom is initially pumped in the level $|c\rangle$, except for spontaneous decay, the time evolution of the atom is

$$|c(t)\rangle = \cos\theta\sin\phi|E_+\rangle e^{-iE_+t} + \cos\theta\cos\phi|E_-\rangle e^{-iE_-t} - \sin\theta|E_0\rangle \qquad (15)$$

We see that $|c\rangle$ evolve $|E_0\rangle$ no oscillating and the oscillation of the states $|\uparrow\rangle$ and $|a\rangle$ due to the coherent superposition of states $|E_\pm\rangle$ just as is determined by Hamiltonian $H'$ in Eq.(14). Not including spontaneous decay, the evolution is entirely coherent, the field oscillates periodically as well. After a period, the atom will come back to $|c\rangle$, the lasing never happens. But including spontaneous decay, there appear two mechanisms influencing the change of the fields. Firstly, it is the oscillation of the states $|\uparrow\rangle$ and $|a\rangle$ due to the coherent superposition of states $|E_\pm\rangle$ which is the same as the theory of lasing with inversion between two-level system: during the interaction of atom and field, the decay of two lower levels $|b\rangle$ and $|c\rangle$ competes with the decay of upper level $|a\rangle$. The former facilitates the amplification of the field, while the latter obstructs the amplification of the field. The probe field will thus be attenuated in the absence of inversion in states $|\uparrow\rangle$ and $|a\rangle$. But there exists another mechanism in favor of the amplification of probe field, which is the generation and decay of dark state $|E_0\rangle$. If $E_\pm - E_0$ is far larger than the decay rate of various energy levels, it might be possible to ignore the off-diagonal elements of the density matrix $\rho_{0+}, \rho_{0-}$ which are very small[16,17]. Thus, the coherence between $|E_0\rangle$ and $|E_\pm\rangle$ is destroyed, the dark stat $|E_0\rangle$ then is generated. $|E_0\rangle$ comes from the coherent transfer $|c\rangle \to |a\rangle \to |b\rangle$, which amplifies the probe field, and become irreversible due to the decay of lower levels $|b\rangle$ and $|c\rangle$. At the same time, what's more important is, during which, the atom doesn't actually arrive at upper level $|a\rangle$, therefore doesn't be influenced by the decay of $|a\rangle$ which obstructs the amplification of the probe field. In order to verify this, we deduce the differential equation of population $\rho_{00}$ of $|E_0\rangle$. For this aim, the atomic model in Fig.2. should be made some reasonable simplifications. The instance that atom decays from $|a\rangle$ to $|b\rangle$ and

$|c\rangle$, or incoherently pumped from $|d\rangle$ to $|b\rangle$ and $|c\rangle$ virtually is a new start that the atom evolve from $|b\rangle$ and $|c\rangle$. So we don't include these process and only seek one process that the atom initially pumped in $|c\rangle$ or $|b\rangle$ evolves coherently as far as decay, thus the decay from $|a\rangle$ to $|b\rangle$ and $|c\rangle$ is only viewed as the decay of $|a\rangle$ itself at the rate $\gamma_a$ which is the sum of $\gamma_{ac}$ and $\gamma_{ab}$. Therefore, from Eq.(3), we obtain

$$\dot{\rho}_{00} = (\gamma_{cd} + \gamma_{bd})\sin\theta\cos\theta \operatorname{Re}\rho_{bc} - \cos^2\theta\gamma_{bd}\rho_{bb} - \sin^2\theta\gamma_{cd}\rho_{cc} \tag{16}$$

It can be seen that the population loss from the dark state $|E_0\rangle$ occurs only due to the decay of states $|b\rangle$ and $|c\rangle$, and is independent of population $\rho_{aa}$ and the decay rate $\gamma_a$. In addition, it also doesn't straightforward be influenced by the detuning $\Delta$. The atom continuously decays from $|a\rangle$ to $|b\rangle$ and $|c\rangle$ or is incoherently pumped from $|d\rangle$ to $|b\rangle$ and $|c\rangle$, during which, the population $\rho_{00}$ of dark state $|E_0\rangle$ restarts the same process, so the consequence of steady state is the summation of these processes. If the state $|c\rangle$ derived from the decay from $|a\rangle$ or the incoherent pumping from $|d\rangle$ is more than $|b\rangle$ derived from the decay from $|a\rangle$ or the incoherent pumping from $|d\rangle$, it is possible to amplify the probe field due to the generation and decay of dark state $|E_0\rangle$. If the amplification effect exceeds the attenuation effect due to the absence of population inversion in states $|a\rangle$ and $|\uparrow\rangle$ in the first mechanism influencing the change of the fields, then the probe field has a net gain. Because state $|E_0\rangle$ doesn't include state $|a\rangle$, to achieve net gain, the condition without population inversion in lower levels $|b\rangle$ and $|c\rangle$ versus upper level $|a\rangle$ is not passive but positive. If the population of $|a\rangle$ increases, it certainly will strengthen the oscillation between $|a\rangle$ and $|\uparrow\rangle$, and attenuate the advantage of gain due to the generation and decay of dark state $|E_0\rangle$. For seeing about the case of population inversion, we seek the basis where the low-level coherence vanishes, i.e. in the representation of the eigenstates of the density submatrix [7], we obtain

$$\begin{pmatrix} \rho_{bb} & \rho_{bc} \\ \rho_{cb} & \rho_{cc} \end{pmatrix}\psi_{1,2} = \lambda_{1,2}\psi_{1,2}; \quad \lambda_{1,2} = (\rho_{bb} + \rho_{cc})/2 \pm \sqrt{(\rho_{bb} - \rho_{cc}/2)^2 + |\rho_{bc}|^2} \tag{17}$$

We compare $\rho_{aa}$ with $\lambda_{1,2}$ and see whether there is population inversion. The response of $\rho_{aa} - \lambda_{1,2}$ versus the detuning $\Delta$ in the units of the decay $\gamma_{bd}$ is plotted in Fig.6(a). We can see that there is not population inversion either between $\rho_{aa}$ and $\lambda_{1,2}$ as is shown in Fig.6(a) or the population inversion between $\rho_{aa}$ and $\rho_{\uparrow\uparrow}$ as is shown in Fig.6(b). From Fig.6(b), we can also see that $\rho_{aa}$ gradually decreases along with the increase of $|\Delta|$, during which, however, there is a peak value of the gain of probe field as is shown in Fig.5(a). It is just the positive effect due to the absence of inversion between the lower levels $|b\rangle$ and $|c\rangle$ versus $|a\rangle$. What should be indicated is that the generation of dark state is essentially a type of the stimulated Raman adiabatic process which is different from the stimulated Raman adiabatic passage that is employed to realize population transfer by means of tuning the fields on and off in counterintuitive order. But it should be a type of stimulated Raman adiabatic process too, because, just as is presented in the previous passage, it is insensitive to the presence of intermediate state $|a\rangle$, hence the decay of $|a\rangle$ and equivalent detuning $\Delta$ from $|a\rangle$.

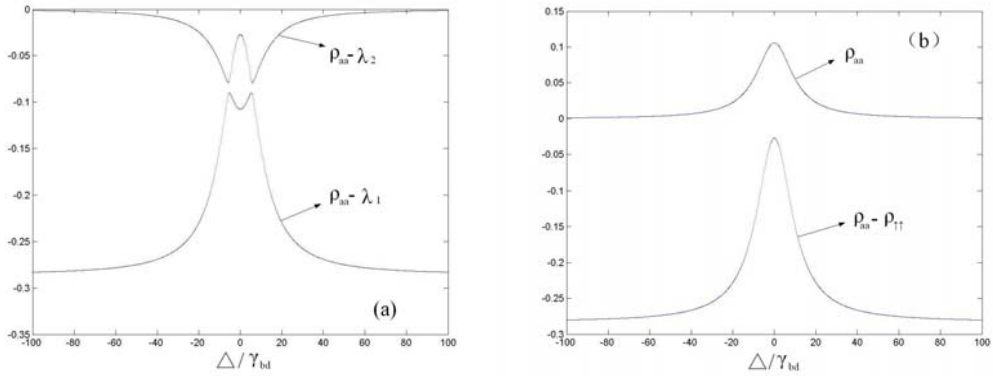

FIG.6. (a) A plot of $\rho_{aa} - \lambda_{1,2}$ versus normalized detuning $\Delta/\gamma_{bd}$, there is population inversion. (b) The response of $\rho_{aa}$ and $\rho_{aa} - \rho_{\uparrow\uparrow}$ versus normalized detuning $\Delta/\gamma_{bd}$. $\rho_{aa}$ gradually decreases along with the increase of $|\Delta|$, and $\rho_{aa} < \rho_{\uparrow\uparrow}$。Values of the parameters are the same as that in fig.5.

In the following, we explain the Fig.5 and the problem represented in Fig.4. For $\Delta_1 = \Delta_2 = \Delta$, with the increase of $|\Delta|$, the rate of oscillation is decelerated, the attenuation effect due to the absence of population inversion in states $|a\rangle$ and $|\uparrow\rangle$ is decelerated too. This can be understood in terms of the time-dependent perturbation theory. The oscillation of $|a\rangle$ and $|\uparrow\rangle$ essentially a type of nonadiabatic stimulated Raman and stimulated Rayleigh processes. Seeing the atomic model in Fig.2, there isn't direct incoherent pumping to $|a\rangle$, thus the firs-order transition rate is only associated with the transition from $|b\rangle$ and $|c\rangle$ to $|a\rangle$. Likewise, there isn't first order transition directly from $|b\rangle$ to $|c\rangle$ or from $|c\rangle$ to $|b\rangle$ in the Hamiltonian $H$ in (1), and the atom is illuminated with the strong coupling filed whose Rabi frequency is larger than the atomic decay rate, then before the atom decays up, it has oscillated many times, the perturbation theory thus is required to retain the higher-order terms. But atom and fields will be influenced by decay and detuning as long as the atomic passes through the upper level $|a\rangle$ in each oscillation. That is to say, it is only necessary to write the expression of first-order and second-order transition rates. Therefore the transition rates of stimulated Raman and stimulated Rayleigh processes[18] read

$$\frac{1}{\tau}(i \to a) \propto \langle a|H_I|i\rangle \delta(\Delta), \quad i = b,c \tag{18a}$$

$$\frac{1}{\tau}(c \to i) \propto \left|\frac{\langle i|H_I|a\rangle\langle a|H_I|c\rangle}{w_1 - w_{ac} + i\gamma_a}\right|^2 \delta(w_1 - w_2 + w_{bc}), \quad i = b,c \tag{18b}$$

$$\frac{1}{\tau}(b \to i) \propto \left|\frac{\langle i|H_I|a\rangle\langle a|H_I|b\rangle}{w_2 - w_{ab} + i\gamma_a}\right|^2 \delta(w_1 - w_2 + w_{bc}), \quad i = b,c \tag{18c}$$

Where $H_I = -(\Omega|a\rangle\langle c| + g|a\rangle\langle b|) + H.c$, $\gamma_a = \gamma_{ac} + \gamma_{ab}$, $\frac{1}{\tau}(c \to i)$ (or $\frac{1}{\tau}(b \to i)$) is the transition rate of $|c\rangle \to |i\rangle$ (or $|b\rangle \to |i\rangle$). From (18a), it can be seen that the first-order transition is remarkably attenuated due to the $\delta$-function. From (18b) and (18c), the second-order transition associated with the nonadiabatic stimulated Raman and stimulated Rayleigh processes is attenuated due to the increase of detuning $\Delta$ ($= w_1 - w_{ac} = w_2 - w_{ab}$), hence the increase of denominators of the right of (18b) and (18c) even at the peak value of $\delta$-function where $w_1 - w_2 + w_{bc} = 0$, i.e., $\Delta_1 = \Delta_2 = \Delta$. On the other hand, just as is discussed in previous passage, because the generation and decay of dark state $|E_0\rangle$ is insensitive to the presence of intermediate state $|a\rangle$, hence the decay of $|a\rangle$ and equivalent detuning $\Delta$ from $|a\rangle$, the increase of detuning $|\Delta|$ doesn't straightforward influence the gain of probe field due to it. So with the increase of $|\Delta|$, the rate of net gain increases as is shown in Fig.5(a). However, $|\Delta|$ ($-|\Delta|$) continuing to increase (decrease), $E_+ - E_0 \to 0$ (or $E_- - E_0 \to 0$), then $|E_+\rangle$ ($|E_-\rangle$) and $|E_0\rangle$ is close to be degenerate, the spectral bandwidth of spontaneous decay is comparable with $E_+ - E_0$ (or $E_- - E_0$), therefore leading to increase of the off-diagonal elements of the density matrix $\rho_{0+}, \rho_{0-}$, the generation and decay of dark state $|E_0\rangle$ is attenuated, then the rate of total net gain gradually decreases as is shown in Fig.5(a). Therefore, it is possible to explain the problem arising in the fig.4 which can not be explained by the Autler-Townes dressed state basis. During the change of $\Delta_1$ from zero to $-20\gamma_{bd}$, $|\Delta_1 - \Delta_2|$ gradually decrease which strengthens the effect of the generation and decay of dark state $|E_0\rangle$, hence the gain of probe field, and there appears a peak value due to the increase of $|\Delta_1|$ as far as $|\Delta_1 - \Delta_2| \to 0$. The change of gain of probe field is consistent with that in Fig5(a).

## V. EXPLAINATION ON RESONANCE

Now we explain the situation on resonance arising in sect. III. From the sixth equation of (5), we obtain:

$$\mathrm{Im}\,\rho_{ba} = \frac{g(\rho_{aa} - \rho_{bb}) - \Omega\,\mathrm{Re}\,\rho_{bc}}{R_{bd} + \gamma_{bd} + \gamma_{ab} + \gamma_{ac}} \qquad (19)$$

For $\rho_{aa} - \rho_{bb} < 0$, seeing Fig.3(b), just because $\mathrm{Re}\,\rho_{bc} < 0$, then it is possible for $\mathrm{Im}\,\rho_{ba} > 0$. For $\mathrm{Re}\,\rho_{bc} < 0$, it might be possible to make the gain of probe due to the generation and decay of dark state $|E_0\rangle$ exceed the attenuation of probe field due to the absence of population inversion in states $|a\rangle$ and $|\uparrow\rangle$ in the first mechanism influencing the change of the fields. With the incoherent pumping between $|c\rangle$ and $|d\rangle$, the atom initially pumped in the level $|c\rangle$, generates dark state $|E_0\rangle$ then decay, thus amplifies the probe field. On the other hand, in the absence of incoherent pumping, the atom in steady state will ultimately focus on ground state $|d\rangle$, the spontaneous decay is no longer active. But with the incoherent pumping, because of the oscillation of the states $|\uparrow\rangle$ and $|a\rangle$, there exists some population of state $|a\rangle$, the spontaneous decay starts to work,. If $\gamma_{ac} > \gamma_{ab}$, and there is the incoherent pumping between $|b\rangle$ and $|d\rangle$ only, the increased population of state $|c\rangle$ due to the coherent transfer $|b\rangle \to |a\rangle \to |c\rangle$ totally comes from the coherent transfer $|b\rangle \to |a\rangle$, so the coherent transfer $|c\rangle \to |a\rangle \to |b\rangle$ is impossible to completely compensate the loss of population of state $|b\rangle$, let alone increase the population of state $|b\rangle$ so as to amplify the probe field. However, if adding the incoherent pumping between $|c\rangle$ and $|d\rangle$, though the average number of thermal photons between them is fewer than that between $|b\rangle$ and $|d\rangle$, but because of $\gamma_{ac} > \gamma_{ab}$, the atom pumped to the level $|c\rangle$ can again decay to $|c\rangle$ from $|a\rangle$, it is possible for the amplification of probe field due to generation and decay of dark state $|E_0\rangle$ to dominate just as is shown in Fig.3. It can also be seen that, for $\gamma_{ac}/\gamma_{ab}$ continuing to increase, the probe field is no longer amplified, it can be attributed to the increase of $\gamma_{ac}$ which weakens the coherence of bare states and makes $\text{Re}\rho_{bc}$ close to zero, and leading to finally that the gain of probe due to the generation and decay of dark state $|E_0\rangle$ does no longer exceed the attenuation of probe field due to the absence of population inversion in states $|a\rangle$ and $|\uparrow\rangle$ in the first mechanism influencing the change of the fields.

## VI. CONCLUSION

The model of four-level atomic system we propose exhibits the lasing without inversion under appropriate conditions. Compared with the three-level system for the lasing without inversion, the four-level atomic system broadens the limits for the real atomic system and thermal radiation for incoherent pumping so as to be more easy to be achieved experimentally, namely no both inversion conditions in the spontaneous decay rate and the average number of thermal photons per mode must be satisfied, but only either one of them is satisfied, furthermore it is no longer to require the incoherent pump between the transition at which lasing occurs, but another lower frequency transition. We find also that when the detuning of probe field is equal to that of coherent coupling field, the rate of gain of probe field can be enhanced. For example, in our model, it is be enhanced about one order. This will facilitate the applications of lasing without inversion. Besides, the presented point of view that there are two mechanisms influencing the change of probe field successful explains various situations in the lasing without inversion, it will help to understand the physics behind the lasing without inversion. The suggest that there are adiabatic and nonadiabatic stimulated Raman processes when a three-level atomic system interacts with two coherent fields needs to be made sure further.

[1]S.E.Harris,Phys.Rev.Lett **62**,1033(1989)
[2]M.O.Scully,S.Y.Zhu,and D.Gavrielides,Phys.Rev.Lett.**62**,2813(1989); S.Y.Zhu,Phys.Rev.A **42,** 5537(1990)
[3] A.Imamoglu,J.E.Field and S.E.Harris , Phys.Rev.Lett .**66** 1154(1991)
[4]YiFu Zhu, Phys.Rev.A **45** R6149(1992)
[5]Olga Kocharovskaya and Paul Mandel, Phys.Rev.A **42**,523(1990) ;Olga Kocharovskaya and Paul Mandel, Phys.Rev.A 45,1997(1992)
[6] YiFu Zhu, Phys.Rev.A **47** 495(1993); YiFu Zhu,Oliver C.Mullins and Min Xiao **47** 602(1993)
[7]Olga Kocharovskaya,Phys.Rep.**219**,175(1992) ; Olga Kocharovskaya,Hyperfine Interactions **107**,187(1997) and M.O.Scully , *ibid*.**219**,191(1992)
[8]E.Arimondo, Phys.Rep.**219**,175(1992)
[9] YiFu Zhu, Phys.Rev.A **55** 4568(1997)

[10] M.O.Scully, 1994 Quantum Opt B**6** 203

[11]G.G.Padmabandu,George R.Welch,Iva N.Shubin *et al*. Phys.Rev.Lett, **76**,2053(1996)

[12].S.Zibrov,M.D.lukin,D.E.Nikonov *et al*. Phys.Rev.Lett, **75**,1499 (1995)

[13].R.Moollow,Phys.Rev.A **5**,2217(1972);F.Y.Wu,S.Ezekiel,M.Ducloy,and B.R.Mollow,Phys.Rev.Lett. **38**, 1077(1977)

[14]G.s.AGARWAL, Phys.Rev.Lett.**67**,980(1991)

[15] M.O.Scully,M.S.Zubairy ,*Quantum Optics*(London ,Cambridge,1997 ),p251

[16] Clude Cohen-Tanudji,Jacques Dupont-Roc, and Gilbert Gryn-berg, *Atom-photon interactions* (wiley,New York,1992),p430

[17] D.Braunstain,R.Shuker, Phys.Rev.A **64** 053812-1(2001)

[18]Rodney Loudon,*The quantum theory of light* (London ,Oxford ,1978),p280